# Antiferromagnetic parametric resonance driven by voltage-controlled magnetic anisotropy

Riccardo Tomasello[1,2], Roman Verba[3], Victor Lopez-Dominguez[1], Francesca Garescì[4], Mario Carpentieri[2], Massimiliano Di Ventra[5], Pedram Khalili Amiri[1,*], Giovanni Finocchio[6,*]

[1]Department of Electrical and Computer Engineering, Northwestern University, Evanston, Illinois 60208, USA

[2]Department of Electrical and Information Engineering, Politecnico of Bari, 70125 Bari, Italy

[3]Institute of Magnetism, Kyiv 03142, Ukraine

[4]Department of Engineering, University of Messina, I-98166, Messina, Italy

[5]Department of Physics, University of California, San Diego, La Jolla, CA 92093, USA

[6]Department of Mathematical and Computer Sciences, Physical Sciences and Earth Sciences, University of Messina, I-98166, Messina, Italy

[*]corresponding authors: gfinocchio@unime.it, pedram@nortwestern.edu

**Abstract**

Voltage controlled magnetic anisotropy (VCMA) is a low-energy alternative to manipulate the ferromagnetic state, which has been recently considered also in antiferromagnets (AFMs). Here, we theoretically demonstrate that VCMA can be used to excite linear and parametric resonant modes in easy-axis AFMs with perpendicular anisotropy, thus opening the way for an efficient electrical control of the Néel vector, and for detection of high-frequency dynamics. Our work leads to two key results: (i) VCMA parametric pumping experiences the so-called "exchange enhancement" of the coupling efficiency and, thus, is 1-2 orders of magnitude more efficient than microwave magnetic fields or spin-orbit-torques, and (ii) it also allows for zero-field parametric resonance, which cannot be achieved by other parametric pumping mechanisms in AFMs with out-of-plane easy axis. Therefore, we demonstrate that the VCMA parametric pumping is the most promising method for coherent excitation and manipulation of AFM order in perpendicular easy-axis AFMs.

# I. INTRODUCTION

Parametric resonance, discovered first in mechanical systems, occurs in various harmonic oscillators in nature, when at least one of their parameters varies periodically in time with an amplitude overcoming a threshold value. In Magnetism, the parametric resonance and other parametric phenomena are very rich, including excitation and amplification of spin waves (SWs), wave front reversal and reversal of momentum relaxation, magnetic soliton compression, and condensation of magnons ([1–5] and references therein). Parametric pumping is usually achieved by external microwave magnetic fields[1,6]. However, in ferromagnets, more energetically-efficient methods involve parametric pumping created by acoustic waves[7,8] or by microwave electric fields via the voltage-controlled magnetic anisotropy (VCMA)[9]. The latter is especially efficient at nanoscale size due to vanishing Ohmic losses[10], and offers a uniquely efficient mechanism of parametric coupling to short (tens of nanometers wavelength) exchange SWs[11].

VCMA has been extensively studied in ferromagnetic metal/dielectric interfaces, typically Fe/MgO or CoFeB/MgO [11–15]. Recently, a very large linear voltage control coefficient $\beta > 0.3$ pJ/Vm [16] has been achieved, which can be increased to $\beta > 1.2$ pJ/Vm [17] in interfaces where ion migration controls the anisotropy. VCMA-driven magnetization dynamics have shown beneficial effects in different applications, such as memory [18,19], magnonic devices [20,21], low-energy motion of skyrmions [22,23], as well as ferromagnetic resonance excitation [14].

Lately, the VCMA effect has been theoretically predicted in antiferromagnets (AFMs) and proposed as an efficient mechanism for the excitation of AFM dynamics[24–28]. In particular, magnetoelectric coefficients as large as $\beta \approx 1.5$ pJ/Vm have been predicted in MgO-capped MnPt films[28]. In addition, indirect evidence of VCMA in AFM thin films was experimentally found in Ref. [29]. AFMs are more abundant in nature than ferromagnets. Recently, they have been receiving a renewed interest thanks to their outstanding properties such as low susceptibility, lack of stray fields, and THz dynamics [30,31]. However, manipulation of AFM order is a more complex task than in ferromagnets due to the high exchange interaction, making it hard to manipulate them by magnetic fields. Recent experiments have demonstrated that the Néel vector dynamics can be driven by electrical currents via spin-orbit torque (SOT) effects [32–38]. This is especially important for the implementation of AFMs in hybrid CMOS/spintronic circuits. VCMA constitutes another viable alternative for electrically driving AFM dynamics, which promises to be more energy-efficient due to reduced Ohmic losses[14,22].

In this work, we demonstrate that VCMA can be successfully applied for the excitation of AFM resonance modes, both in the linear and parametric resonance regimes. Our results point out that parametric resonance in AFMs is very efficient for the excitation of large-amplitude precession of the Néel vector. Although parametric resonance in AFMs driven by microwave magnetic fields has been known for fifty years [39–42], here we show that the mechanism of parametric coupling of VCMA drive to AFM resonance modes is completely different from the one of microwave magnetic field. Therefore, VCMA parametric pumping is not only 1-2 orders of magnitude more efficient than microwave magnetic field pumping, but also allows for parametric excitation in zero bias magnetic field, which is impossible to achieve with magnetic field pumping. Our results open the possibility of realization of power-efficient high frequency AFM-based devices, such as tunable electrical detectors [43,44].

The paper is organized as follows. In Sec. II, theoretical analysis of VCMA-driven AFM dynamics is performed. Section III presents results of micromagnetic simulations of AFM dynamics under VCMA drive and comparison with theoretical prediction. In Sec. IV, we consider the efficiency of VCMA drive by comparing it to alternative excitation mechanisms, i.e. microwave magnetic field, and SOT. Finally, conclusions are made in Sec. V.

## II. THEORY

### A. Model and basic equations

We study the magnetization dynamics of a thin AFM nanoelement with out-of-plane (OOP) uniaxial magnetic anisotropy, as AFMs with OOP anisotropy were predicted to demonstrate VCMA. In both analytical calculations and micromagnetic simulations, we use a continuous two-sublattice model of an AFM, which rigorously applies for many AFMs, and is one of the most comprehensive model of AFM dynamics[45]. In particular, this model has been already used to describe AFM-based THz oscillators [46], detectors [44], and soliton dynamics [47–49]. The model describes the antiferromagnetic order by considering two sublattices characterized by a normalized magnetization vectors $\boldsymbol{m}_j = \boldsymbol{M}_j / M_s$, $j = 1,2$ ($M_s$ is the saturation magnetization of the two sublattices $M_{s1} = M_{s2} = M_s$), which dynamics is governed by two coupled Landau-Lifshitz-Gilbert (LLG) equations[46,47]

$$\frac{\partial \boldsymbol{m}_j}{\partial t} = -\gamma_0 \boldsymbol{m}_j \times \boldsymbol{B}_{\text{eff},j} + \alpha_G \boldsymbol{m}_j \times \frac{\partial \boldsymbol{m}_j}{\partial t}, \qquad (1)$$

where $\gamma_0$ is the gyromagnetic ratio, $\alpha_G$ is the Gilbert damping parameter, and $\boldsymbol{B}_{\text{eff},j}$ is the effective field acting on $j$-th sublattice, which consists of the exchange (homogeneous intersublattice AFM interaction and nonuniform intrasublattice exchange), uniaxial perpendicular anisotropy, VCMA, magneto-dipolar $B_{\text{dip}}$, as well as external field contributions.

Here, we are concentrating on the VCMA contribution. When an electric field $E_{ac}$ is applied to the AFM surface - using a gate separated by a dielectric layer, see Sec. III below - at a microwave frequency $f_{ac}$, the perpendicular magnetic anisotropy is modulated with the same frequency. The resulting effective anisotropy field acting on $j$-th sublattice, thus, becomes

$$\boldsymbol{B}_{a,j} = \left[ B_{a,0} + \Delta B_{VCMA} \sin\left(2\pi f_{ac} t + \phi_{VCMA}\right) \right] \left(\boldsymbol{m}_j \cdot \boldsymbol{e}_z\right) \boldsymbol{e}_z, \quad (2)$$

where $B_{a,0} = 2K_{u,0}/M_s$, $K_{u,0}$ is the uniaxial anisotropy constant at zero applied voltage, $\Delta B_{VCMA} = 2\beta E_{ac}/(t_{AFM} M_s)$ is the anisotropy field modulation due to VCMA, with $\beta$ being the magnetoelectric coefficient (VCMA efficiency), and $t_{AFM}$ being the AFM film thickness, and $\phi_{VCMA}$ is the phase of the applied microwave voltage signal. It is worth noting that the VCMA-related effective field is proportional both to the external voltage, and to the OOP ($z$) component of the sublattice magnetization (in general, time-dependent). This aspect makes the effect of the VCMA completely different from the microwave magnetic field effect, as shown in the following.

## B. Coupling of VCMA drive to AFM eigenmodes

The expression of the VCMA-related effective field in Eq. (2) is used for the analysis of the VCMA action on AFM dynamic modes. Here, we restrict the analysis to the case of spatially-uniform AFM dynamics, i.e. we consider the case of AFM resonance, which is typically achieved in submicron thin AFM elements, as confirmed below by micromagnetic simulations. The VCMA interaction with propagating SWs or spatially-nonuniform AFM modes can be carried out within the same formalism.

We assume that the AFM can be biased by an in-plane magnetic field $B_x$, applied, for definiteness, in the $x$ direction. Under such a field, the static magnetization $\boldsymbol{\mu}_j = \boldsymbol{m}_{j,0}$ of the sublattices tilts, so that $\boldsymbol{\mu}_1 = \left(\sin\phi, 0, \cos\phi\right)$, $\boldsymbol{\mu}_2 = \left(\sin\phi, 0, -\cos\phi\right)$, where $\sin\phi = B_x / \left(B_{a,0} + 2B_{ex}\right)$. Here, $B_{ex}$ is the effective fields of the homogeneous AFM exchange, which is the strongest effective field in an AFM, $B_{ex} \gg B_a, B_{dip}$. Excitations of an AFM with uniaxial anisotropy are well-known. The excitation spectrum consists of 2 linear modes, lower frequency Mode1 and higher frequency Mode 2. Their frequencies, when accounting for the demagnetization fields of thin AFM nanoelement, are[1,4,50]:

$$\omega_{(1)}^2 = \omega_a \left(\omega_a + 2\omega_{ex}\right) - \frac{\omega_H^2 \omega_a}{\omega_a + 2\omega_{ex}} \quad (3)$$

$$\omega_{(2)}^2 = \omega_a \left(\omega_a + 2\omega_{ex}\right) + \omega_H^2 \frac{2\omega_{ex} - \omega_a + 2\omega_M}{\omega_a + 2\omega_{ex}} \quad (4)$$

where $\omega_a = \gamma_0 B_a$, $\omega_{ex} = \gamma_0 B_{ex}$, $\omega_H = \gamma_0 B_x$, $\omega_M = \gamma_0 \mu_0 M_s$. At low and moderate fields ($B_x \ll 2B_{ex}$), the frequencies of linear modes can be approximated to [1,51]:

$$\omega_{(1)}^2 \approx 2\omega_a \omega_{ex} - \frac{\omega_H^2 \omega_a}{2\omega_{ex}}; \quad \omega_{(2)}^2 \approx 2\omega_a \omega_{ex} + \omega_H^2 \tag{5}$$

The frequency of Mode 1 is slightly dependent (decreasing) on the field, while the frequency of Mode 2 increases with $B_x$ (examples are shown below in Sec. III). At zero field, the modes are, naturally, degenerated. Comparing to ferromagnets, where the linear mode has a frequency of the order of $\omega_{FM} \sim \omega_a$ (in an unbiased case), the frequencies of the AFM modes are $\sqrt{2\omega_{ex}/\omega_a} \gg 1$ times larger, which constitutes the so-called effect of "exchange enhancement" of AFM dynamic characteristics, caused by huge effective fields of homogeneous AFM exchange[45,51]. Many, but not all AFM characteristics experiences this enhancement.

The structure of the AFM modes are fully characterized by a net dynamic magnetization $\boldsymbol{m}_{(\nu)} = (\boldsymbol{m}_{1,(\nu)} + \boldsymbol{m}_{2,(\nu)})/2$, and a dynamic Neel vector $\boldsymbol{l}_{(\nu)} = (\boldsymbol{m}_{1,(\nu)} - \boldsymbol{m}_{2,(\nu)})/2$, where $\nu = 1,2$ is the mode index (in order not to mix mode index and sublattice index, here and in the following the mode index is in the brackets). The net magnetic moment and dynamic AFM vector of the lower mode (Mode 1) are:

$$\boldsymbol{m}_1 = \frac{1}{\sqrt{A_1}} \begin{pmatrix} \cos\phi \\ 0 \\ 0 \end{pmatrix}, \quad \boldsymbol{l}_1 = \frac{1}{\sqrt{A_1}} \begin{pmatrix} 0 \\ i\cos\phi\sqrt{\omega_a + 2\omega_{ex}/\omega_a} \\ -\sin\phi \end{pmatrix}, \tag{6}$$

where the coefficient $A_1 = 4\omega_1/\omega_a$ is chosen so that the norm of spin-wave modes $i\sum_j \boldsymbol{m}_{j,(\nu)}^* \cdot \boldsymbol{\mu}_j \times \boldsymbol{m}_{j,(\nu)} = 1$ [50]. This mode has only in-plane net magnetization $m_x$, and, thus, can linearly couple to in-plane microwave magnetic field $b_x$ only, while the dynamic Neel vector is characterized by nonzero $y$ and $z$ components (the latter, however, is present only under a nonzero bias field and is, typically, weak). The AFM vector of the upper mode, in contrast, is $x$-polarized, while the net magnetization has $y$ and $z$ components:

$$\boldsymbol{m}_2 = \frac{1}{\sqrt{A_2}} \begin{pmatrix} 0 \\ -i\omega_2/\omega_a + 2\omega_{ex} \\ \sin\phi \end{pmatrix}, \quad \boldsymbol{l}_2 = \frac{1}{\sqrt{A_2}} \begin{pmatrix} -\cos\phi \\ 0 \\ 0 \end{pmatrix}, \tag{7}$$

with $A_2 = 4\omega_2/\omega_a + 2\omega_{ex}$.

The coupling of the VCMA to the AFM eigenmodes is convenient to consider within the framework of SW perturbation theory [50]. Using the expression for the VCMA effective field in Eq.

(1), one finds, in a general case, that the linear coupling to the VCMA is proportional to the OOP dynamic component of the AFM vector $l_z$. Therefore, there is no coupling to the Mode 2, and the coupling to the Mode 1 appears only at a nonzero bias field and is proportional to $\sin\phi$. The fact that the linear coupling is defined by $l_z$ and not by $m_z$ - as the coupling with the external field -, underlines a crucial difference in the action of the VCMA on the AFM dynamics compared to the magnetic field action, in which it is also the key element to excite parametric dynamics (see below). Indeed, the linear components (at the excitation frequency) of the VCMA effective field are opposite in the two sublattices $\boldsymbol{B}_{\text{eff},j} = \Delta B_{\text{VCMA}} \mu_{z,j} \boldsymbol{e}_z$ because of the opposite orientations of the OOP static magnetization components, hence it could excite modes with anti-phase OOP dynamic magnetization. In contrast, an external magnetic field, obviously, acts in phase on both sublattices.

The effective driving field acting on the Mode 1 is equal to

$$\tilde{b} = -\frac{\Delta B_{\text{VCMA}} \sin\phi \cos\phi}{2\sqrt{\omega_1/\omega_a}}, \tag{8}$$

and corresponding excited mode amplitude is

$$c_1 = \frac{\gamma \tilde{b}}{i\,\omega_{ac} - \omega_1 + \Gamma_1}, \tag{9}$$

which is related with the amplitude of Neel vector oscillations as $l_{\max} = 2|c_\nu|\,l_\nu\sqrt{1-|c_\nu|^2/4}$. In Eq. (9), $\Gamma$ is the damping rate, which is given by $\Gamma_{(\nu)} = \alpha_G \varepsilon_{(\nu)} \omega_{(\nu)}$ with

$$\varepsilon_1 = \frac{\omega_1^2 + \omega_a^2}{2\omega_1 \omega_a}, \qquad \varepsilon_2 = \frac{\left(\omega_a + 2\omega_{ex}\right)^2 + \omega_2^2}{2\omega_2\left(\omega_a + 2\omega_{ex}\right)}. \tag{10}$$

being the coefficients dependent on the precession ellipticity of the modes[50]. In the range of small bias fields $\varepsilon_1 \approx \varepsilon_2 \approx \sqrt{\omega_{ex}/2\omega_a}$, we can approximate the damping rate to $\Gamma_1 = \Gamma_2 \approx \alpha_G \omega_{ex}$. It is clear that the damping rate also exhibits the effect of exchange enhancement, leading to wide AFM resonance curves.

The enhanced damping rate results in a weak efficiency of the linear excitation of the AFM modes. The coupling efficiency in Eq. (8) shows no exchange enhancement, and it is also proportional to small values of $\sin\phi$, and $\sqrt{\omega_a/\omega_1} \approx 1/\sqrt[4]{2\omega_{ex}/\omega_a}$. Therefore, the amplitude of the excited mode, given by the coupling efficiency and damping, is small. We can conclude that the VCMA is a good mechanism for the linear excitation of AFM resonance modes. Nevertheless, in the case of sufficient bias magnetic field, the VCMA could be the most efficient mechanism for the excitation of the Mode 1. This aspect is promising from an experimental point of view because the VCMA can provide a larger dynamical contribution to the effective field and low parasitic losses simultaneously. The same problem of small excitation rate stands true for other linear excitation mechanisms, especially for microwave field, as discussed in Sec. IV.

Now, we consider the parametric coupling of the VCMA to the AFM modes. The calculation of the parametric interaction efficiency due to the VCMA pumping is done within the same SW perturbation formalism[50]. In a general case, we obtain

$$V_1 = \frac{i\gamma}{A_1}\left[\left(\frac{2\omega_{ex}}{\omega_a}\cos^2\phi - \sin^2\phi\right)\cos^2\phi + \sin^2\phi\right],$$
$$V_2 = \frac{i\gamma}{A_2}\left[-\left(1 - \frac{\omega_{(2)}^2}{(\omega_a + 2\omega_{ex})^2}\right)\cos^2\phi + \sin^2\phi\right]$$
(11)

The parametric coupling coefficients $V$ are defined as usual, so that the parametric term in the equation of motion of the mode amplitude is derived as $dc_{(v)}/dt + ... = V_{(v)} b_p c_{(v)}^* e^{-i\omega_p t}$, where $b_p = \Delta B_{VCMA}$ is the pumping amplitude, and $\omega_p$ is the VCMA pumping frequency. In the range of small bias fields, expressions (11) are reduced to:

$$V_1 \approx i\gamma \frac{\cos^3\phi}{4}\sqrt{\frac{2\omega_{ex}}{\omega_a}}, \quad V_2 \approx -i\gamma \frac{\cos 2\phi}{4}\sqrt{\frac{2\omega_{ex}}{\omega_a}},$$
(12)

These expressions clearly demonstrate that the parametric coupling to VCMA pumping in small bias fields also exhibit the exchange enhancement, as it follows from the multiplier $\sqrt{2\omega_{ex}/\omega_a}$. This exchange enhancement is related to strong ellipticity of the magnetization precession, which creates a large longitudinal component at the double oscillation frequency, responsible for the parametric coupling.

The exchange enhancement of the parametric coupling to VCMA is a key result of our theoretical calculations. This enhancement retains the parametric excitation mechanism efficient despite the exchange enhanced damping rates. Indeed, the threshold of the parametric excitation is determined by the ratio $b_{th} = V/\Gamma$, and the exchange enhancement of the coupling efficiency "compensates" the one of the damping rate, resulting in accessible values of the excitation threshold. Below, by means of micromagnetic simulations, we also show that, above the threshold, the amplitudes of parametrically-excited modes grows fast, and can reach large values up to $c \sim 1$.

With the increase of the bias field, the efficiency of ellipticity-caused parametric coupling decreases, as it is usual for the so-called "tilted" parametric pumping[1,2], and vanishes at $\phi \to \pi/2$ (this contribution is described by the first term in Eq. (11)). Another contribution to the parametric coupling efficiency ($\sin^2\phi$ term in Eq. (11)) is specific for anisotropy-driven parametric pumping[8,20], and therefore for VCMA, and it increases with the static magnetization tilt. However, this contribution does not exhibit the exchange enhancement, and the VCMA driven parametric excitation becomes much less efficient at $\phi \to \pi/2$.

### III. MICROMAGNETIC SIMULATIONS

#### A. Micromagnetic model

In order to verify our theoretical predictions, we performed systematic micromagnetic simulations of the AFM dynamics using an in-house numerical framework[46,47]. In addition, micromagnetic simulations allow us to study overthreshold dynamics, find stationary amplitudes of parametrically excited modes, and compare excitation efficiency by VCMA with other possible driving forces. Last but not least aim is to validate the above used approximation of spatially-uniform dynamics, and excitation of linear eigenmodes of the AFM sample. Indeed, AFMs are characterized by nonlinear dynamics even in small samples, as proved by excitations of magnetic solitons and domain walls [45,51]. Therefore, it is not straightforward that the parametric excitation gives rise to uniform AFM modes.

The magnetization dynamics of an AFM nanoelement with OOP uniaxial magnetic anisotropy, which can be also biased by an in-plane (IP) magnetic field $B_x$, is studied, and different driving forces – magnetic fields, SOT [32,34,35,52], as well as VCMA – are considered and compared. A sketch of the studied structure is shown in Fig. 1. It is a 5-terminal device where a squared 200x200 nm$^2$ AFM voltage gate [53,54] is placed on top of a Heavy Metal (HM) cross bar. The four terminals of the cross bar allow for injecting an ac electrical current $j_{HM}(t) = J_{HM} \sin(2\pi f_{ac} t + \phi_{HM})$ into the Platinum HM, thus generating the SOT with spin-polarization $p_y$ ($p_x$) along the $y$-direction ($x$-direction) when the current flows between the terminals A-A' (B-B'). The fifth terminal is used to apply an ac voltage $V(t) = V \sin(2\pi f_{ac} t + \phi_{VCMA})$ (Fig. 1(b)), which modulates the OOP anisotropy $K_u(t) = K_{u,0} + \Delta K_u \sin(2\pi f_{ac} t + \phi_{VCMA})$ of the AFM because of the VCMA effect. Also, a microwave magnetic field $b(t)$ with different polarizations could be applied to the structure. The proposed structure can be used for experimental investigation of the VCMA and SOT-driven AFM dynamics in the same sample, as well as charting an experimentally-viable pathway towards practical implementation of the discussed phenomena.

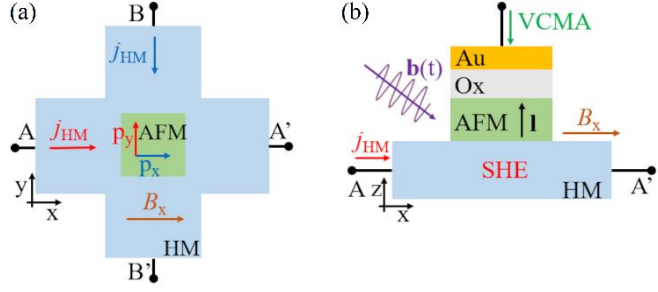

FIG.1 A sketch of the studied structure, (a) -the top view of the AFM/Pt cross bar bilayer, (b) - the *x-z* view of the AFM voltage gate on top of the Pt cross bar.

The micromagnetic simulations are based on the above mentioned two-sublattice continuous model of AFM in Eq. (1). The exchange effective field acting on the 1$^{st}$ sublattice include three terms:

$$\mathbf{B}_{1,\text{ex}} = \frac{2A_{11}}{M_s}\nabla^2 \mathbf{m}_1 + \frac{4A_0}{a^2 M_s}\mathbf{m}_2 + \frac{A_{12}}{M_s}\nabla^2 \mathbf{m}_2, \quad (13)$$

and the expression for $\mathbf{B}_{2,\text{ex}}$ is achieved by the index permutation $1 \leftrightarrow 2$. Here, $a$ is the lattice constant, $A_{11} > 0$ and $A_{22} > 0$ are the inhomogeneous intra-lattice contributions (assumed to be equal, $A_{11} = A_{22}$), $A_{12} < 0$ is the inhomogeneous inter-sublattice contribution, which is neglected in this study, and $A_0 < 0$, is the homogeneous inter-sublattice contribution to the exchange energy. The effect of SOT from the spin-Hall effect (SHE) is described by addition of the following torque [35] to Eq. (1):

$$\mathbf{T}_j = d_J \left( \frac{\theta_{i-DLT} j_{HM}(t)}{t_{AFM}} \right) (\mathbf{m}_j \times \mathbf{m}_j \times \mathbf{p}) \quad (14)$$

with $d_J$ being the torque coefficient given by $d_J = \frac{g\mu_B}{2eM_S}$, where $g$ is the Landè factor, $\mu_B$ is the Bohr magneton, and $e$ is the electron charge. The coefficient $\theta_{i-DLT}$ takes into account the efficiency of the charge/spin current conversion of the current $j_{HM}(t) = J_{HM} \sin(2\pi f_{ac} t + \phi_{HM})$ flowing in the HM. The vector $\mathbf{p}$ is the direction of the spin-polarization. The physical parameters in the simulations are similar to previous studies on metallic AFMs, such as PtMn, FeMn and FeRh, namely $M_s = 566$ kA/m, $A_{11} = A_{22} = 5.0$ pJ/m, $A_0 = -0.248$ pJ/m, $K_u = 28.3$ kJ/m$^3$, $a$=0.5 nm, $\theta_{i-DLT} = 0.15$, and $\alpha = 0.02$ [35]. The thickness of AFM layer is $t_{AFM} = 1$ nm. The cell size in the simulation was set to 4 x 4 x 1 nm$^3$.

### B. VCMA-driven AFM dynamics

First, we performed characterization of linear eigenmodes by exciting them with different drives (details on which mode is excited by microwave field or SOT with a given polarization are discussed in Sec. IV). The frequencies of the simulated linear modes (which, obviously, does not depend on the type of linear drive) coincide with analytical calculations – in the range of small and moderate bias fields, the frequency of Mode 1 is weakly field-dependent, while the frequency of Mode 2 increases with the field (Fig. 2). In the following, if other is not stated, the external filed is fixed to $B_x$ = 1800 mT in order to easily distinguish between the modes.

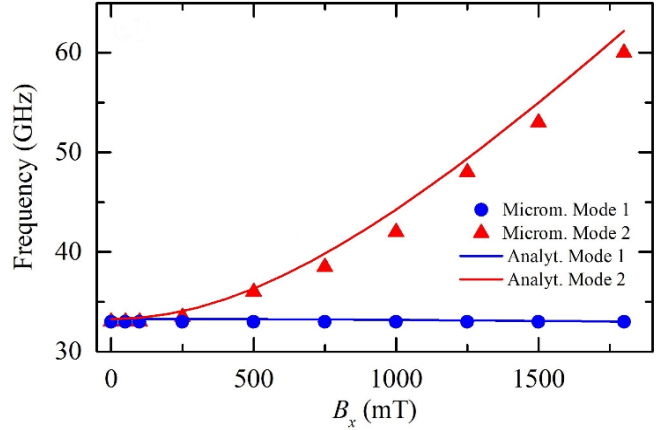

FIG.2 Resonance frequency of the Mode 1 and 2 versus external bias field. The solid lines represent the analytical results from Eqs. (3-4), the symbols are related to the micromagnetic results

Now, we focus on the VCMA-driven excitation of the AFM resonance modes. The simulated resonance curves under the VCMA drive of different amplitude are shown in Fig. 3. At small amplitudes of VCMA drive ($B_{VCMA}$ < 45 mT), the resonance curve exhibits a broad resonance peak at the frequency of 33 GHz (Fig. 3(a)), which is the Mode 1 eigenfrequency. In contrast, we find no linear response of the Mode 2 to the VCMA drive. This behaviour is in full accordance with our theoretical calculations, which shows that the linear VCMA couples only to the Mode 1. The amplitude of the linearly-excited Mode 1 nicely follows the calculated linear dependence in Eq. (9). However, even applying a large VCMA drive of $\Delta B_{VCMA}$ = 80 mT leads to very small oscillations amplitude of the linearly-excited mode $l_y \approx 0.08$.

At larger VCMA drive amplitudes, $B_{VCMA} \geq 45$ mT, the simulations show the appearance of a second peak, which is located at about the double frequency of Mode 1. The amplitude of this peak abruptly increases with VCMA beyond a threshold value (see Figs. 3(a) – (c)). These characteristics are a clear evidence of the parametric resonance. Importantly, the parametrically-driven dynamics of the AFM nanoelement remains almost perfectly spatially-uniform (in all the studied range of VCMA drives, as well as at different bias magnetic fields,), thus validating our theoretical approach.

The theoretically-calculated parametric excitation threshold for $B_x$ = 1800 mT is $\Delta B_{VCMA,th}$ = 48.6 mT, which agrees well with the simulation result $\Delta B_{VCMA,th} \approx 48$ mT. If we consider the predicted VCMA magnetoelectric coefficient of $\beta \approx 1.5$ pJ/Vm [28], and an MgO thickness of 2 nm in the structure, the required voltage to achieve the parametric resonance is about 17 mV, which should be not hard to access in experiments and utilize for various applications. At higher excitation frequency and higher VCMA drive (central frequency 124.4 GHz and the threshold of 96.5 mT), we also observe parametric resonance of the Mode 2 in the simulations (not shown), which also correlates well with theoretical predictions.

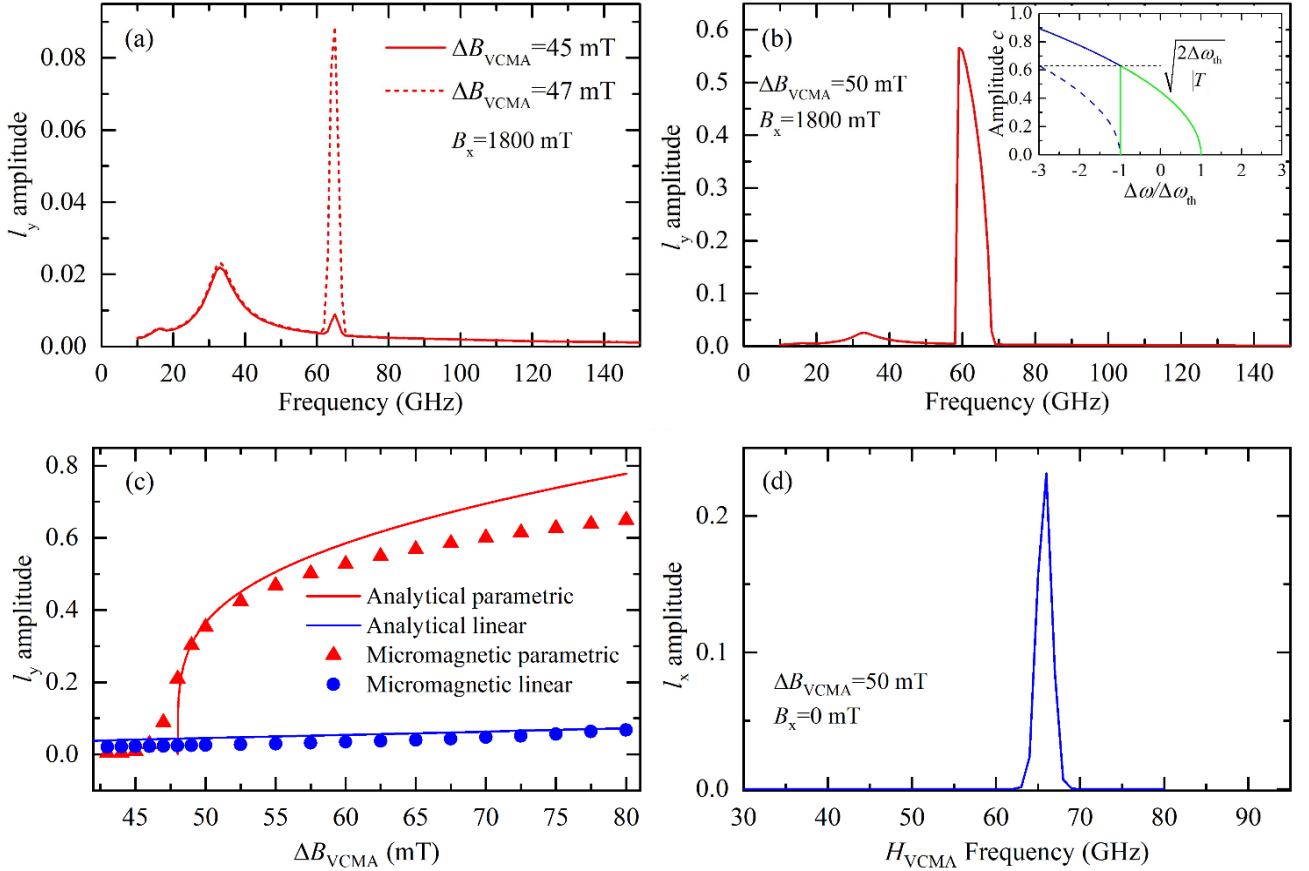

FIG. 3 (a, b) Resonant response of the Néel vector y-component for different values of the VCMA drive at $B_x = 1800$ mT; inset in (b) shows theoretical shape of the parametric resonance peak (Eq. (15)) – typically, the peak has the green-curve-like shape, solid blue curve can be accessed in specific cases (see text). (c) Amplitude of the Néel vector oscillations excited by linear (33 GHz) and parametric (65 GHz) VCMA drive. (d). Parametric excitation of the Néel vector x-component when the VCMA drive of 50 mT is applied to an unbiased AFM.

Just above the parametric excitation threshold, the resonance peak is narrow, especially in comparison with linear resonance (Fig. 3(a)). With the increase of the parametric pumping, the peak becomes wider and acquires a characteristic antisymmetric "triangle-like" shape (Fig. 3(b)). To describe the peak shape as well as finding the amplitudes of the parametrically-excited modes, we need to consider nonlinear effects, which limit the growth of the parametric instability. In a confined geometry with discrete SW spectrum, typically, the most important nonlinear effect is the nonlinear frequency shift of the $v$-th mode frequency $\omega_{(v)} = \omega_{(v),0} + T_{(v)} \left| c_{(v)} \right|^2$, accounting for which, the amplitude $c_{(v)}$ of the parametrically-excited mode is calculated as[9]:

$$\left| c_{(v)} \right|^2 = \frac{\Delta \omega}{T_{(v)}} + \frac{\sqrt{\left| V_{(v)} \Delta B_{\mathrm{VCMA}} \right|^2 - \Gamma_{(v)}^2}}{\left| T_{(v)} \right|}, \tag{15}$$

where $\Delta \omega = \omega_{\mathrm{p}}/2 - \omega_{(v),0}$ is the detuning from the exact parametric resonance. Although the calculation of the nonlinear frequency shift is a complex task, one can use the following trick in the case of the low-frequency mode. The relation between both static and dynamic magnetization components of the sublattices is the same ($m_{x,1} = m_{x,2}$, $m_{y,1} = -m_{y,2}$, $m_{z,1} = -m_{z,2}$), which allows for reducing the two coupled LLG equations for the sublattices into one effective equation, and apply ready-to-use equations based on the Hamiltonian formalism [55,56]. In this way, we find $T_1 \approx -\omega_{ex}/2$ (for $B_x \ll 2B_{ex}$). In fact, the excited mode amplitude given by Eq. (15) not always can be reached. In the case of low thermal noise, the SW mode is excited within the frequency range $\left| \Delta \omega \right| < \Delta \omega_{\mathrm{th}} = \sqrt{\left| V_{(v)} \Delta B_{\mathrm{VCMA}} \right|^2 - \Gamma_{(v)}^2}$, in which small-amplitude SWs become unstable due to the parametric pumping. In this case, the parametric resonance curve has a characteristic "triangular" shape (see green curve in the inset of Fig. 3(b)) with the maximum at the left (right) edge for negative (positive) nonlinear frequency shift equal to $\left| c_{(v),\max} \right|^2 = 2\sqrt{\left| V_{(v)} \Delta B_{\mathrm{VCMA}} \right|^2 - \Gamma_{(v)}^2} / \left| T_{(v)} \right|$. This result reproduces well the micromagnetic outcomes (Fig. 3(c)). Some discrepancy at large VCMA drive is common and related with the utilization of Taylor expansion in the Hamiltonian formalism, which becomes less accurate at large precession amplitudes (typically, for $l_y > 0.5$). However, if thermal fluctuations are large (overcoming the dashed curve in the inset of Fig. 3(b)), or the excitation frequency is continuously swept from the right, one can access the part of the curve beyond $-\Delta \omega_{\mathrm{th}}$. The maximal frequency detuning and peak amplitude in this more complex case are determined by other nonlinear mechanisms (nonlinearity of parametric interaction efficiency or/and nonlinearity of damping), consideration of which lies beyond the scope of this work.

## IV. COMPARISON WITH ALTERNATIVE DRIVES

Above, we found that, by means of VCMA parametric pumping, it is possible to excite large-amplitude SW modes in an AFM nanoelement. Let us briefly consider alternative mechanism and the efficiency of spin wave excitation by their means, starting in this subsection from the linear excitation.

The oldest and well-established mechanism involves the use of microwave magnetic fields. Depending on the polarization of the magnetic field, it can excite either Mode1 or Mode 2, or both modes, which is determined by the net dynamic magnetization of the modes $\boldsymbol{m}_{(\nu)}$. Thus, according to Eqs. (6) and (7), the microwave field $\boldsymbol{b}_x$ excites only Mode 1, while fields $\boldsymbol{b}_y$ and $\boldsymbol{b}_z$ excite Mode 2 (the last one only in the presence of static bias field $B_x$), in full accordance with micromagnetic data. The coupling efficiency, however, shows the same problem as the linear coupling with VCMA drive – they are inversely proportional to $\sqrt[4]{2\omega_{ex}/\omega_a}$ and, accounting for the enhanced damping, the excited SW amplitudes at the experimentally-achievable microwave drive are very low. The exception is OOP microwave field. The coupling to it exhibits a "partial exchange enhancement" $\tilde{b} \sim \sqrt[4]{2\omega_{ex}/\omega_a}$, being, however, proportional to another weak value $\sin\phi$. Thus, microwave magnetic field is not a choice for the linear excitation of OOP easy-axis AFM independently of the polarization.

Much recent approach to excite AFM dynamics is the application of SOT, e.g., by means of SHE. In the simulations, Mode 1 (2) was observed when a SOT with a spin-polarization parallel (perpendicular) to the external field is applied and the dynamic AFM vector has $y$ ($x$)-component of the Néel vector $l_{y\,(x)}$ at the excitation frequency. Exemplary resonance curves of SOT-driven AFM dynamics are shown in the inset of Fig. 4, additionally underlining large width of the resonance curves in the linear excitation regime. The mode selectivity is also easy to find theoretically considering the mode structure and SOT effective field.

Figure 4 shows the Néel vector oscillation amplitude excited by SOT. Here, the ac electrical current is chosen large but achievable in experiment – it is sufficiently small to prevent non-magnetic phenomena, such as electromigration [38], but is not far from this limit. The amplitude of Mode 1 is larger than the one of Mode 2 and is weakly dependent on the field in the studied range. Contrarily, the amplitude of Mode 2 decreases with the field. Overall, the maximum achievable oscillation amplitude is reasonable $l_{max} \sim 0.075$, and significantly larger than ones achievable with microwave magnetic field excitation. Nevertheless, such oscillation level can be insufficient for some applications, e.g., for application of these devices as detectors, considering that the magnetization precession is at not large amplitude and the electrical readout mechanisms for AFM order, developed so far, are not very efficient. Also, large precession amplitudes are indispensable for devices based on nonlinear SW interactions.

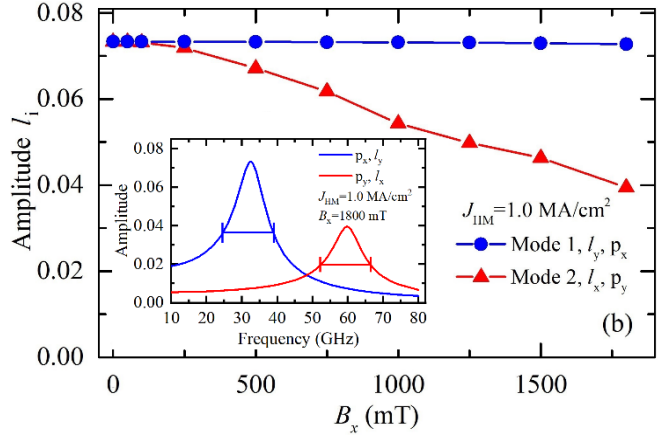

FIG.4 Micromagnetic results for the amplitude of the Mode 1 and 2 ($l_y$ or $l_x$, respectively) as a function of external bias field when a SOT with $p_y$ or $p_x$ is applied. Inset shows exemplary resonance curves at $B_x = 1800$ mT. Driving current density in HM layer is $J_{HM}=1.0$ MA/cm$^2$.

We have already mentioned above that the coupling efficiency of SOT to AFM modes experiences "partial exchange enhancement", being proportional to $\tilde{b} \sim \sqrt[4]{2\omega_{ex}/\omega_a}$. It is much better than the microwave field excitation, but not enough to fully compensate the exchange enhancement of the damping. Thus, SOT in AFM is less efficient for linear SW excitation than in ferromagnets. In a general case, field-like SOT can be also present in AFMs, but it does not lead to any significant differences in the excitation efficiency, as is discussed in the Appendix A.

Now, let us look on the parametric excitation. First, it is worth noting that microwave field-driven parametric resonance in AFMs is generally well-known. However, almost all the previous works considered easy-plane AFMs [40–42], which demonstrate good parametric coupling with magnetic field pumping. Easy-axis AFMs, considered in this work, show a completely different behavior. The parametric coupling with OOP magnetic field $b_z$ is identically zero since it affects the sublattices with opposite phases. The parametric resonance is possible only in a biased state ($B_x >0$) under "parallel pumping" $\boldsymbol{b}_p=b_x\boldsymbol{e}_x$. The efficiency of the parametric coupling to this pumping is easy to calculate within the same approach, as for VCMA, which yields

$$V_\nu = -i\gamma \frac{\varepsilon_\nu \sin\phi}{2}, \qquad (16)$$

where ellipticity-related coefficient $\varepsilon_\nu$ is given by Eq. (10). Although this coupling efficiency also demonstrates exchange enhancement ($\varepsilon_\nu \sim \sqrt{2\omega_{ex}/\omega_a}$), it is also proportional to $\sin\phi$, and, consequently, at low and moderate bias fields, is small. In particular, it is much smaller than the parametric coupling for the VCMA pumping, as shown in Fig. 5. In addition, this figure clearly shows that the VCMA-driven parametric excitation is achievable in zero bias magnetic field, as also confirmed by the micromagnetic simulations (see Fig. 3(d)). This is important both from a fundamental and practical point of view, since this behavior cannot be obtained by any other here-studied means, and underlines one more time the crucial difference between the VCMA action on AFM materials compared to the field action.

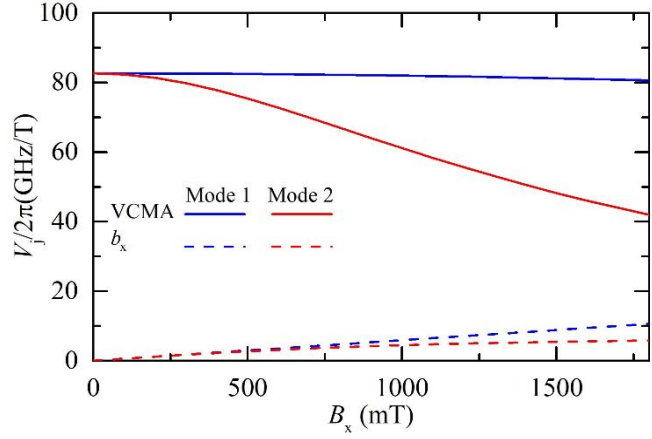

FIG. 5 Parametric interaction efficiency of the VCMA pumping and magnetic field pumping with both AFM resonance modes.

Finally, let us consider the possibility of parametric resonance excitation by SOT. The effective field of SOT is given by $b_{\text{SOT},j} = b_{\text{SOT}}(m_j \times p) = b_{\text{SOT}}\left[(c_\nu m_{j,(\nu)} \times p) + (c_\nu^* m_{j,(\nu)}^* \times p) + (\mu_j \times p)\right]$. Within the framework of perturbation theory[50], the coefficients, which describe parametric coupling, are $b_{(\nu)}$ for the part of perturbation field, linear in $c_\nu^*$, and $\tilde{S}_{\nu,\nu}$ in calculation of which only the static part of perturbation field is accounted for (see Eqs. (2.4-2.5) in[50]). Both of them are identically zero independently of the spin current polarization, since $b_{(\nu)} \sim \sum_j m_{j,(\nu)}^* \cdot \left(m_{j,(\nu)}^* \times p_j\right) \equiv 0$ and $\tilde{S}_{(\nu),(\nu)} \sim \sum_j \mu_j \cdot \left(\mu_j \times p_j\right) \equiv 0$. Thus, degenerate parametric process cannot be driven by SOT from SHE. Nondegenerate process (splitting of pumping into 2 different modes, $\omega_P \to \omega_{(1)} + \omega_{(2)}$) is also impossible – its efficiency is given by $\sum_j m_{j,(1)}^* \cdot \left(m_{j,(2)}^* \times p_j\right) \equiv 0$ and one can easily check that contributions from different sublattices compensate each other.

To summarize, we put all the considered cases in Table I, that shows the correspondence between drive and excited AFM mode, as well as how the coupling relates to the exchange enhancement rate $\xi = \sqrt{2\omega_{\text{ex}}/\omega_{\text{a}}}$. One can see that, among all the considered mechanisms, SOT is the most suitable for *linear* excitation of AFM eigenmodes. The most efficient mechanism is, however, the VCMA-driven parametric resonance, as it is the only mechanism exhibiting a "full exchange enhancement" $V \sim \sqrt{2\omega_{ex}/\omega_a}$ both for unbiased and weakly biased AFMs. It is worth noting that the exchange enhancement rate is, indeed, large and is of a principal importance for the AFM dynamics. For the studied AFM, this ratio is $\sqrt{2\omega_{ex}/\omega_a} = 11.84$, and could reach even higher values for other AFMs with stronger homogeneous AFM exchange interaction. In addition to high "magnetic efficiency", the VCMA drive has a perfect electric efficiency in terms of low Joule heating losses and other parasitic losses. Overall, the VCMA parametric pumping results to be the most promising method for coherent excitation and manipulation of AFM order in easy-axis AFMs with OOP anisotropy.

| Drive Type | Coupling efficiency | | Parametric coupling, $V$ |
|---|---|---|---|
| | Linear coupling, $\tilde{b}$ | | |
| | **Mode 1** | **Mode 2** | **Both modes** |
| SOT with spin-polarization *perpendicular* to $B_x$ | 0 | $\sim \sqrt{\xi}$ | 0 |
| SOT with spin-polarization *parallel* to $B_x$ | $\sim \sqrt{\xi}$ | 0 | 0 |
| Microwave field $b_x$ | $\sim 1/\sqrt{\xi}$ | 0 | $\sim \xi \sin\phi$ |
| Microwave field $b_y$ | 0 | $\sim 1/\sqrt{\xi}$ | 0 |
| Microwave field $b_z$ | 0 | $\sim \sqrt{\xi} \sin\phi$ | 0 |
| VCMA | $\sim \sin\phi/\sqrt{\xi}$ | 0 | $\sim \xi$ |

TABLE I. Summary of the excited modes of an AFM with OOP easy axis for different excitation sources. The order of the coupling efficiency is also indicated, where $\xi = \sqrt{2\omega_{ex}/\omega_a} \gg 1$ is the "exchange enhancement ratio". The proportionality to $\sin\phi$ underlines that the excitation mechanism can only work for an AFM biased by an external magnetic field.

## V. CONCLUSIONS

In summary, we have analyzed, by means of micromagnetic simulations and analytical theory, the excitation of resonant modes in a uniaxial perpendicular AFM comparing different excitation source: magnetic fields, SOT, and VCMA. The linear excitation can be achieved by all the sources, where the particular excited mode depends on combination of field/SOT polarization, and bias magnetic field. However, amplitudes of SW modes, which could be reached in an experiment, are not large and does not exceed $l_i \sim 0.05$-$0.1$, because of the exchange enhancement of the damping rate, which cannot be completely balanced by any linear drive source.

In contrast, VCMA parametric pumping demonstrates exchange enhancement of the coupling rate, allowing, thus, for the excitation of SW modes with unprecedentedly large precession amplitude. The parametric resonance could in principle be excited by microwave magnetic fields, however, the parametric interaction efficiency for the field pumping is much lower than the one for VCMA pumping. In addition, the VCMA advantageously allows for parametric excitation even at zero magnetic field.

Compared to SOT, which can only excite linear modes, the VCMA stands as an unique electrical pumping source for efficient excitation of large-amplitude coherent dynamics in easy-axis AFMs. This is a key and promising result for AFMs device applications, which should not be based on linear modes but on parametric excitation.

## ACKNOWLEDGEMENT

This work was supported under the Grant 2019-1-U.0. ("Diodi spintronici rad-hard ad elevata sensitività - DIOSPIN") funded by the Italian Space Agency (ASI) within the call "Nuove idee per la componentistica spaziale del futuro", and the project PRIN 2020LWPKH7 funded by the Italian Ministry of University and Research. R. V. acknowledges support by the Ministry of Education and Science of Ukraine (project # 0121U110090). This work was also supported by a grant from the U.S. National Science Foundation, Division of Electrical, Communications and Cyber Systems (NSF ECCS-1853879), and by the National Science Foundation Materials Research Science and Engineering Center at Northwestern University (NSF DMR-1720319). MD acknowledges support from the DOE under Grant No. DESC0020892. The work of R.T. at the Northwestern University was sponsored by the Petaspin Association (www.petaspin.com).

## APPENDIX A: EFFECT OF THE FIELD-LIKE TORQUE

We simulated the effect of the SO-field-like torque (FLT) $B_{FLT}$. We performed systematic micromagnetic simulations to study how $B_{FLT}$ affects the Néel vector dynamics acting simultaneously with the damping-like torque-related field (DLT) $B_{DLT} = \left[2\mu_0 d_j \theta_{i-DLT} j_{HM}(t)\right] / \gamma_0 t_{AFM}$, where $d_J = \frac{g\mu_B}{2eM_S}$, with $g$ being the Landè factor, $\mu_B$ the Bohr magneton, $e$ the electron charge, and $M_S$ the saturation magnetization. $\mu_0$ is the vacuum permeability, $\theta_{i-DLT}$ takes into account the efficiency of the charge/spin current conversion of the current $j_{HM}(t)$ flowing in the heavy metal, $\gamma_0$ is the gyromagnetic ratio, and $t_{AFM}$ is the AFM thickness.

In Fig. 6, we summarize the results (the time evolution of the spatially-averaged Néel vector components) achieved at the resonance frequency, where the larger effect is observed, for spin-polarization $p_x$ and $p_y$ and three different bias fields $B_x$. We consider four values of the ratio $B_{FLT}/B_{DLT}$ respectively equal to 0.0, 0.2, 0.4, and 0.6 (which we also refer to as FLT0.0, FLT0.2, FLT0.4, and FLT0.6). The FLT is considered along the x- (y-) direction according to the direction of the spin-polarization $p_y$ ($p_x$), and the two torques are perpendicular in the sample plane.

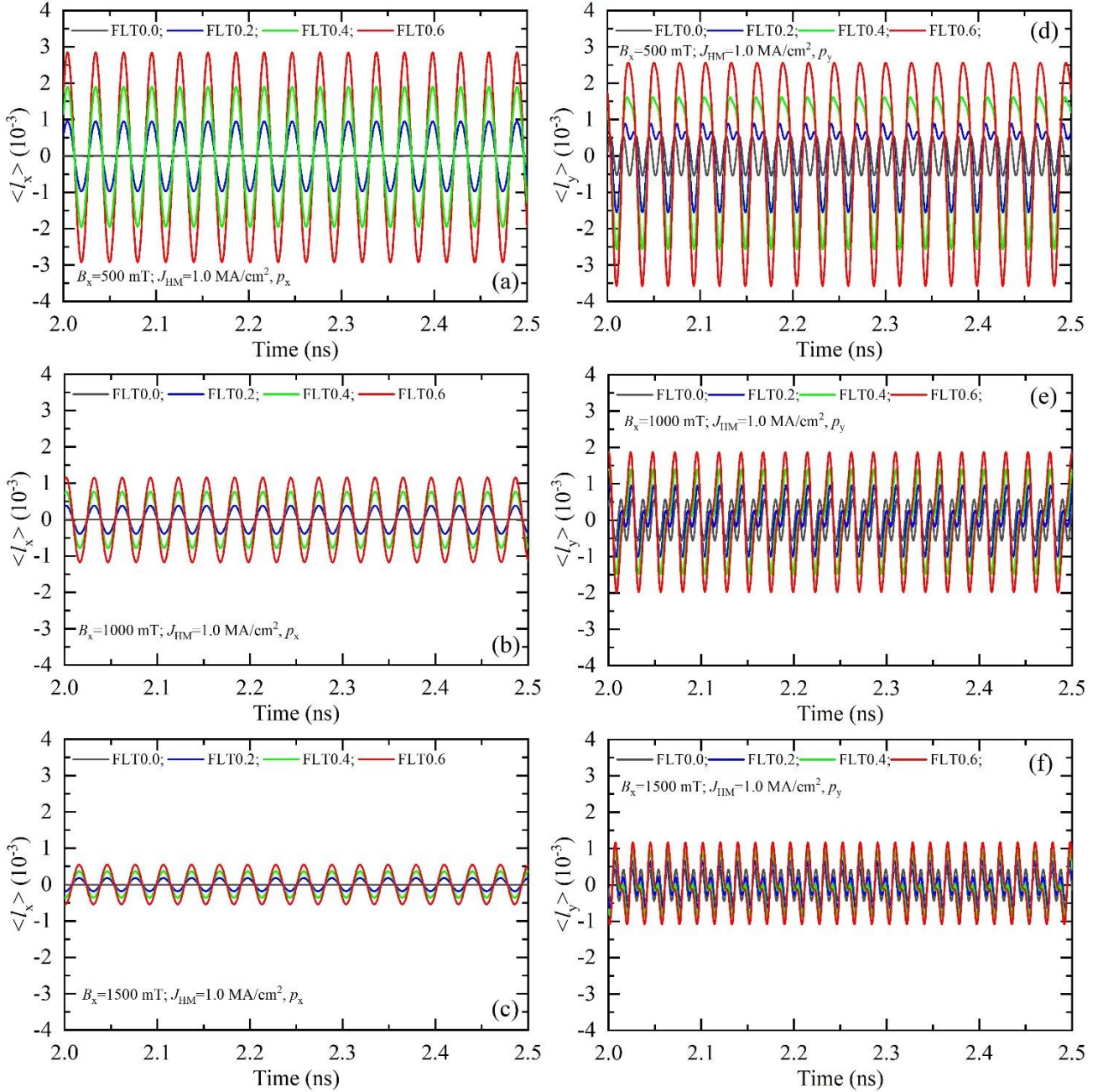

FIG. 6 Time evolution of the spatially-averaged Néel vector components for four values of the $B_{FLT}/B_{DLT}$ ratio equal to 0.0, 0.2, 0.4, and 0.6 at the corresponding resonant frequency. (a) – (c) x-component and (d) – (f) y-component as a function of $B_x$=500, 1000, and 1500 mT respectively.

For spin-polarization $p_x$, the y- and z-components do not exhibit significant changes under the FLT. On the other hand, the x-component $\langle l_x \rangle$ oscillation amplitude increases as a function of the FLT strength for each $B_x$, but such an increase is smaller as $B_x$ gets larger (see also Fig. 7(a)).

For spin-polarization $p_y$, the x- and z-components do not exhibit significant changes under the FLT. Whereas, the y-component $\langle l_y \rangle$ not only increases its oscillation amplitude (see also Fig. 7(b)), but also changes its oscillation frequency (Fig 6(d) – (f)). Specifically, at zero FLT, the DLT promotes an oscillation at twice the input frequency, while a non-zero FLT leads the oscillation frequency to be at the same frequency as the input one. Therefore, we observe a trade-off between these two effects, which yields the existence of a threshold value of the FLT for each $B_x$. Below such a threshold, $\langle l_y \rangle$ is characterized by those two modes simultaneously, and beyond the threshold, $\langle l_y \rangle$ oscillates at the input frequency. For instance, we can compare the blue curve in Fig. 6(d) for FLT<0.3DLT (double mode) with the grey curve at zero FLT (single mode at twice the input frequency), and with the green and red curves at higher FLT (single mode at the input frequency).

The different behavior for different spin-polarizations is ascribed to the relative orientation between the ac field responsible for the FLT and the bias field $B_x$. In the first case ($p_x$), they are perpendicular to each other and the only effect is the increase of the oscillation amplitude. In the second case ($p_y$), they are parallel to each other and both amplitude and frequency change.

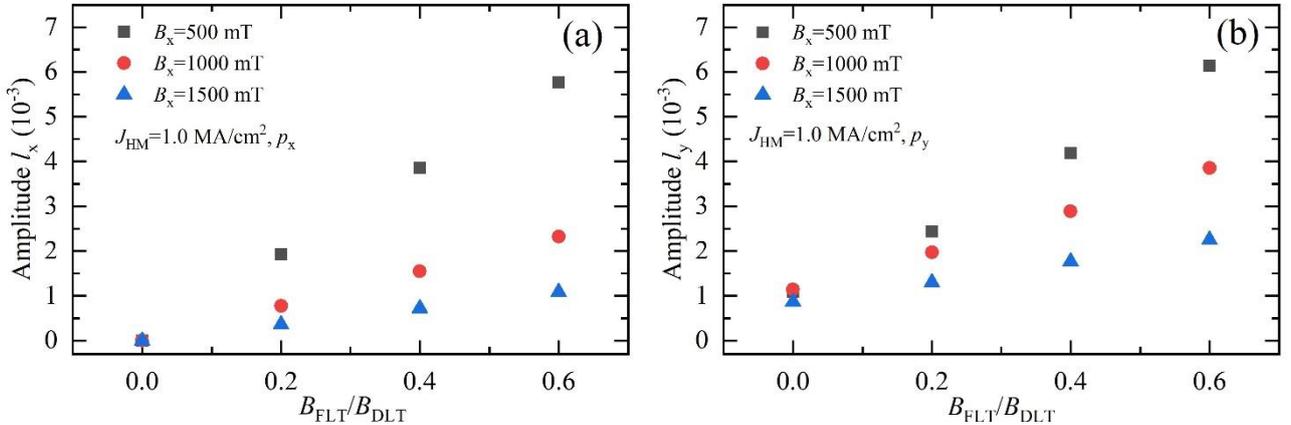

FIG. 7 Amplitude of the Néel vector x- (a) and y-(b) component as a function of the FLT and for three values of the external field at the corresponding resonance frequency.

**REFERENCES**

[1] G. Gurevich and A. Melkov, Magnetization Oscillations and Waves, *Magnetization Oscillations and Waves* (CRC Press, New York, 1996).

[2] V.S. L'vov, Wave Turbulence Under Parametric Excitation, *Wave Turbulence Under Parametric*

*Excitation* (Springer Berlin Heidelberg, Berlin, Heidelberg, 1994).

[3] O. V. Prokopenko, D.A. Bozhko, V.S. Tyberkevych, A. V. Chumak, V.I. Vasyuchka, A.A. Serga, O. Dzyapko, R. V. Verba, A. V. Talalaevskij, D. V. Slobodianiuk, Y. V. Kobljanskyj, V.A. Moiseienko, S. V. Sholom, and V.Y. Malyshev, Recent Trends in Microwave Magnetism and Superconductivity, Ukr. J. Phys. **64**, 888 (2019).

[4] T. Brächer, P. Pirro, and B. Hillebrands, Parallel pumping for magnon spintronics: Amplification and manipulation of magnon spin currents on the micron-scale, Phys. Rep. **699**, 1 (2017).

[5] O.R. Sulymenko, O. V. Prokopenko, V.S. Tyberkevych, A.N. Slavin, and A.A. Serga, Bullets and droplets: Two-dimensional spin-wave solitons in modern magnonics, Low Temp. Phys. **44**, 602 (2018).

[6] S. Urazhdin, V. Tiberkevich, and A. Slavin, Parametric excitation of a magnetic nanocontact by a microwave field, Phys. Rev. Lett. **105**, 237204 (2010).

[7] I. Lisenkov, A. Jander, and P. Dhagat, Magnetoelastic parametric instabilities of localized spin waves induced by traveling elastic waves, Phys. Rev. B **99**, 184433 (2019).

[8] M. Geilen, R. Verba, A. Nicoloiu, D. Narducci, A. Dinescu, M. Ender, M. Mohseni, F. Ciubotaru, M. Weiler, A. Müller, B. Hillebrands, C. Adelmann, and P. Pirro, Parametric Excitation and Instabilities of Spin Waves driven by Surface Acoustic Waves, arxiv:2201.04033 (2022).

[9] Y.-J. Chen, H.K. Lee, R. Verba, J.A. Katine, I. Barsukov, V. Tiberkevich, J.Q. Xiao, A.N. Slavin, and I.N. Krivorotov, Parametric Resonance of Magnetization Excited by Electric Field, Nano Lett. **17**, 572 (2017).

[10] Q. Wang, M. Kewenig, M. Schneider, R. Verba, F. Kohl, B. Heinz, M. Geilen, M. Mohseni, B. Lägel, F. Ciubotaru, C. Adelmann, C. Dubs, S.D. Cotofana, O. V. Dobrovolskiy, T. Brächer, P. Pirro, and A. V. Chumak, A magnonic directional coupler for integrated magnonic half-adders, Nat. Electron. **3**, 765 (2020).

[11] T. Maruyama, Y. Shiota, T. Nozaki, K. Ohta, N. Toda, M. Mizuguchi, A.A. Tulapurkar, T. Shinjo, M. Shiraishi, S. Mizukami, Y. Ando, and Y. Suzuki, Large voltage-induced magnetic anisotropy change in a few atomic layers of iron, Nat. Nanotechnol. **4**, 158 (2009).

[12] S. Ikeda, K. Miura, H. Yamamoto, K. Mizunuma, H.D. Gan, M. Endo, S. Kanai, J. Hayakawa, F. Matsukura, and H. Ohno, A perpendicular-anisotropy CoFeB-MgO magnetic tunnel junction., Nat. Mater. **9**, 721 (2010).

[13] P. Khalili Amiri, Z.M. Zeng, J. Langer, H. Zhao, G. Rowlands, Y.J. Chen, I.N. Krivorotov, J.P. Wang, H.W. Jiang, J.A. Katine, Y. Huai, K. Galatsis, and K.L. Wang, Switching current reduction using perpendicular anisotropy in CoFeB-MgO magnetic tunnel junctions, Appl. Phys. Lett. **98**, 112507 (2011).

[14] J. Zhu, J.A. Katine, G.E. Rowlands, Y.-J. Chen, Z. Duan, J.G. Alzate, P. Upadhyaya, J. Langer, P.K. Amiri, K.L. Wang, and I.N. Krivorotov, Voltage-Induced Ferromagnetic Resonance in Magnetic Tunnel Junctions, Phys. Rev. Lett. **108**, 197203 (2012).

[15] P.K. AMIRI and K.L. WANG, Voltage-Controlled Magnetic Anisotropy In Spintronic Devices, SPIN **02**, 1240002 (2012).

[16] T. Nozaki, A. Kozioł-Rachwał, M. Tsujikawa, Y. Shiota, X. Xu, T. Ohkubo, T. Tsukahara, S. Miwa, M. Suzuki, S. Tamaru, H. Kubota, A. Fukushima, K. Hono, M. Shirai, Y. Suzuki, and S. Yuasa, Highly efficient voltage control of spin and enhanced interfacial perpendicular magnetic anisotropy in iridium-doped Fe/MgO magnetic tunnel junctions, NPG Asia Mater. **9**, e451 (2017).

[17] A. Fassatoui, J.P. Garcia, L. Ranno, J. Vogel, A. Bernand-Mantel, H. Béa, S. Pizzini, and S. Pizzini, Reversible and Irreversible Voltage Manipulation of Interfacial Magnetic Anisotropy in Pt/Co/Oxide Multilayers, Phys. Rev. Appl. **14**, 064041 (2020).

[18] W.-G. Wang, M. Li, S. Hageman, and C.L. Chien, Electric-field-assisted switching in magnetic tunnel junctions, Nat. Mater. **11**, 64 (2012).

[19] C. Grezes, F. Ebrahimi, J.G. Alzate, X. Cai, J.A. Katine, J. Langer, B. Ocker, P. Khalili Amiri, and K.L. Wang, Ultra-low switching energy and scaling in electric-field-controlled nanoscale magnetic tunnel junctions with high resistance-area product, Appl. Phys. Lett. **108**, 012403 (2016).

[20] R. Verba, M. Carpentieri, G. Finocchio, V. Tiberkevich, and A. Slavin, Excitation of Spin Waves in an In-Plane-Magnetized Ferromagnetic Nanowire Using Voltage-Controlled Magnetic Anisotropy, Phys. Rev. Appl. **7**, 064023 (2017).

[21] Q. Wang, A. V. Chumak, L. Jin, H. Zhang, B. Hillebrands, and Z. Zhong, Voltage-controlled nanoscale reconfigurable magnonic crystal, Phys. Rev. B **95**, 134433 (2017).

[22] W. Kang, Y. Huang, C. Zheng, W. Lv, N. Lei, Y. Zhang, X. Zhang, Y. Zhou, and W. Zhao, Voltage Controlled Magnetic Skyrmion Motion for Racetrack Memory, Sci. Rep. **6**, 23164 (2016).

[23] X. Wang, W.L. Gan, J.C. Martinez, F.N. Tan, M.B.A. Jalil, and W.S. Lew, Efficient skyrmion transport mediated by a voltage controlled magnetic anisotropy gradient, Nanoscale **10**, 733 (2018).

[24] G. Zheng, S.-H. Ke, M. Miao, J. Kim, R. Ramesh, and N. Kioussis, Electric field control of magnetization direction across the antiferromagnetic to ferromagnetic transition, Sci. Rep. **7**, 5366 (2017).

[25] V. Lopez-Dominguez, H. Almasi, and P.K. Amiri, Picosecond Electric-Field-Induced Switching of Antiferromagnets, Phys. Rev. Appl. **11**, 024019 (2019).

[26] P.A. Popov, A.R. Safin, A. Kirilyuk, S.A. Nikitov, I. Lisenkov, V. Tyberkevich, and A. Slavin, Voltage-Controlled Anisotropy and Current-Induced Magnetization Dynamics in Antiferromagnetic-Piezoelectric Layered Heterostructures, Phys. Rev. Appl. **13**, 044080 (2020).

[27] Y. Su, M. Li, J. Zhang, J. Hong, and L. You, Voltage-controlled magnetic anisotropy in antiferromagnetic L10-MnPt and MnPd thin films, J. Magn. Magn. Mater. **505**, 166758 (2020).

[28] P.-H. Chang, W. Fang, T. Ozaki, and K.D. Belashchenko, Voltage-controlled magnetic anisotropy in antiferromagnetic MgO-capped MnPt films, Phys. Rev. Mater. **5**, 054406 (2021).

[29] Y. Wang, X. Zhou, C. Song, Y. Yan, S. Zhou, G. Wang, C. Chen, F. Zeng, and F. Pan, Electrical Control of the Exchange Spring in Antiferromagnetic Metals, Adv. Mater. **27**, 3196 (2015).

[30] T. Jungwirth, X. Marti, P. Wadley, and J. Wunderlich, Antiferromagnetic spintronics, Nat. Nanotechnol. **11**, 231 (2016).

[31] V. Baltz, A. Manchon, M. Tsoi, T. Moriyama, T. Ono, and Y. Tserkovnyak, Antiferromagnetic spintronics, Rev. Mod. Phys. **90**, 015005 (2018).

[32] P. Wadley, B. Howells, J. Elezny, C. Andrews, V. Hills, R.P. Campion, V. Novak, K. Olejnik, F. Maccherozzi, S.S. Dhesi, S.Y. Martin, T. Wagner, J. Wunderlich, F. Freimuth, Y. Mokrousov, J. Kune, J.S. Chauhan, M.J. Grzybowski, A.W. Rushforth, K.W. Edmonds, B.L. Gallagher, and T. Jungwirth, Electrical switching of an antiferromagnet, Science **351**, 587 (2016).

[33] M.J. Grzybowski, P. Wadley, K.W. Edmonds, R. Beardsley, V. Hills, R.P. Campion, B.L. Gallagher, J.S. Chauhan, V. Novak, T. Jungwirth, F. Maccherozzi, and S.S. Dhesi, Imaging Current-Induced Switching of Antiferromagnetic Domains in CuMnAs, Phys. Rev. Lett. **118**, 057701 (2017).

[34] J. Godinho, H. Reichlová, D. Kriegner, V. Novák, K. Olejník, Z. Kašpar, Z. Šobáň, P. Wadley, R.P. Campion, R.M. Otxoa, P.E. Roy, J. Železný, T. Jungwirth, and J. Wunderlich, Electrically induced and detected Néel vector reversal in a collinear antiferromagnet, Nat. Commun. **9**, 4686 (2018).

[35] J. Shi, V. Lopez-Dominguez, F. Garesci, C. Wang, H. Almasi, M. Grayson, G. Finocchio, and P. Khalili Amiri, Electrical manipulation of the magnetic order in antiferromagnetic PtMn pillars, Nat. Electron. **3**, 92 (2020).

[36] S. DuttaGupta, A. Kurenkov, O.A. Tretiakov, G. Krishnaswamy, G. Sala, V. Krizakova, F. Maccherozzi, S.S. Dhesi, P. Gambardella, S. Fukami, and H. Ohno, Spin-orbit torque switching of an antiferromagnetic metallic heterostructure, Nat. Commun. **11**, 5715 (2020).

[37] H. Tsai, T. Higo, K. Kondou, T. Nomoto, A. Sakai, A. Kobayashi, T. Nakano, K. Yakushiji, R. Arita, S. Miwa, Y. Otani, and S. Nakatsuji, Electrical manipulation of a topological antiferromagnetic state, Nature **580**, 608 (2020).

[38] S. Arpaci, V. Lopez-Dominguez, J. Shi, L. Sánchez-Tejerina, F. Garesci, C. Wang, X. Yan, V.K. Sangwan, M.A. Grayson, M.C. Hersam, G. Finocchio, and P. Khalili Amiri, Observation of current-induced switching in non-collinear antiferromagnetic IrMn3 by differential voltage measurements, Nat. Commun. **12**, 3828 (2021).

[39] M.H. Seavey, Nuclear and Electronic Spin-Wave Relaxation Rates in the Hexagonal Antiferromagnet CsMnF 3, J. Appl. Phys. **40**, 1597 (1969).

[40] L.A. Prozorova and A.S. Borovik-Romanov, Parametric excitation of spin waves inantiferromagnetic CsMnF3, JETP Lett. **10**, 201 (1969).

[41] B.Y. Kotyuzhanskii and L.A. Prozorova, Spin waves relaxation in antiferromagnetic CsMnF3, JETP Lett. **38**, 1233 (1974).

[42] J. Barak, S.M. Rezende, A.R. King, and V. Jaccarino, Parallel-pumping studies of magnon damping in MnF2, Phys. Rev. B **21**, 3015 (1980).

[43] O. Gomonay, T. Jungwirth, and J. Sinova, Narrow-band tunable terahertz detector in antiferromagnets via staggered-field and antidamping torques, Phys. Rev. B **98**, 104430 (2018).

[44] A. Safin, V. Puliafito, M. Carpentieri, G. Finocchio, S. Nikitov, P. Stremoukhov, A. Kirilyuk, V. Tyberkevych, and A. Slavin, Electrically tunable detector of THz-frequency signals based on an antiferromagnet, Appl. Phys. Lett. **117**, 222411 (2020).

[45] E.G. Galkina and B.A. Ivanov, Phenomenological description of spin dynamics in antiferromagnets: Short history and modern development, Low Temp. Phys. **47**, 765 (2021).

[46] V. Puliafito, R. Khymyn, M. Carpentieri, B. Azzerboni, V. Tiberkevich, A. Slavin, and G. Finocchio, Micromagnetic modeling of terahertz oscillations in an antiferromagnetic material driven by the spin Hall effect, Phys. Rev. B **99**, 024405 (2019).

[47] L. Sánchez-Tejerina, V. Puliafito, P. Khalili Amiri, M. Carpentieri, and G. Finocchio, Dynamics of domain-wall motion driven by spin-orbit torque in antiferromagnets, Phys. Rev. B **101**, 014433 (2020).

[48] A. Salimath, F. Zhuo, R. Tomasello, G. Finocchio, and A. Manchon, Controlling the deformation of antiferromagnetic skyrmions in the high-velocity regime, Phys. Rev. B **101**, 024429 (2020).

[49] R. Tomasello, L. Sanchez-Tejerina, V. Lopez-Dominguez, F. Garescì, A. Giordano, M. Carpentieri, P.K. Amiri, and G. Finocchio, Domain periodicity in an easy-plane antiferromagnet with Dzyaloshinskii-Moriya interaction, Phys. Rev. B **102**, 224432 (2020).

[50] R. Verba, V. Tiberkevich, and A. Slavin, Damping of linear spin-wave modes in magnetic nanostructures: Local, nonlocal, and coordinate-dependent damping, Phys. Rev. B **98**, 104408 (2018).

[51] L.R. Walker, Spin Waves. A. I. Akhiezer, V. G. Bar'yakhtar, and S. V. Peletminskii. Translated from the Russian by S. Chomet. S. Doniach, translation editor. North-Hol-land, Amsterdam; Interscience (Wiley), New York, 1968. Science **163**, 923 (1969).

[52] S.Y. Bodnar, L. Šmejkal, I. Turek, T. Jungwirth, O. Gomonay, J. Sinova, A.A. Sapozhnik, H.-J. Elmers, M. Kläui, and M. Jourdan, Writing and reading antiferromagnetic Mn2Au by Néel spin-

orbit torques and large anisotropic magnetoresistance, Nat. Commun. **9**, 348 (2018).

[53] B.G. Park, J. Wunderlich, X. Martí, V. Holý, Y. Kurosaki, M. Yamada, H. Yamamoto, A. Nishide, J. Hayakawa, H. Takahashi, A.B. Shick, and T. Jungwirth, A spin-valve-like magnetoresistance of an antiferromagnet-based tunnel junction, Nat. Mater. **10**, 347 (2011).

[54] O.R. Sulymenko, O. V. Prokopenko, V.S. Tyberkevych, and A.N. Slavin, Terahertz-Frequency Signal Source Based on an Antiferromagnetic Tunnel Junction, IEEE Magn. Lett. **9**, 3104605 (2018).

[55] P. Krivosik and C.E. Patton, Hamiltonian formulation of nonlinear spin-wave dynamics: Theory and applications, Phys. Rev. B **82**, 184428 (2010).

[56] R. Verba, V. Tiberkevich, and A. Slavin, Hamiltonian formalism for nonlinear spin wave dynamics under antisymmetric interactions: Application to Dzyaloshinskii-Moriya interaction, Phys. Rev. B **99**, 174431 (2019).